\documentclass[aps,prd,showpacs,floatfix,preprint]{revtex4}
\usepackage{amsmath,bm}
\usepackage{graphicx}
\usepackage{epsfig}
\begin{document}

\title{Effect of hyperon-hyperon interaction on bulk viscosity and
r-mode instability in neutron stars}
\author{Debarati Chatterjee and Debades Bandyopadhyay}
\affiliation{Saha Institute of Nuclear Physics, 1/AF Bidhannagar, 
Kolkata-700064, India}

\begin{abstract}
We investigate the effect of hyperon matter including hyperon-hyperon 
interaction on bulk viscosity. Equations of state are constructed within the 
framework of a relativistic field theoretical model where baryon-baryon 
interaction is mediated by the exchange of scalar and vector mesons. 
Hyperon-hyperon interaction is also taken into account by the exchange of two 
strange mesons. This interaction results in a smaller maximum mass neutron star
compared with the case without the interaction. The coefficient of bulk 
viscosity due to the non-leptonic weak process is determined by these equations
of state. The interacting hyperon matter reduces the bulk viscosity coefficient
in a neutron star interior compared with the no interaction case. The r-mode 
instability is more effectively suppressed in hyperon-hyperon interaction case 
than that without the interaction.
\end{abstract}
\maketitle

\section{Introduction}

Neutron stars may undergo various instabilities which  
are associated with unstable modes of oscillation \cite{And01}. 
To understand oscillatory 
modes of a neutron star one has to study its stability properties. Each 
physical restoring force acting on a fluid element leads to a new family of 
oscillatory modes. Pulsation modes of neutron stars are classified according to
different restoring forces. Here we are interested in Coriolis restored 
inertial r-modes. It was shown that r-modes were subject to 
Chandrasekhar-Friedman-Schutz gravitational radiation instability in rapidly
rotating neutron stars \cite{And01,And98,Fri98,Lin98,And99,Ster}. This 
instability may play an important role in regulating spins of young neutron 
stars as well as old, accreting neutron stars in low mass x-ray binaries 
(LMXBs) and provides a plausible explanation for the absence of very fast 
rotating neutron stars in nature. This finds support from the study of eleven 
nuclear-powered 
millisecond pulsars whose spin frequencies are known from burst oscillations 
has shown that the spin distribution has an upper limit at
730 Hz \cite{Cha,Cha2}. Recently observed fastest rotating neutron star is a 
radio pulsar with a spin frequency 716 Hz \cite{Hes}. The r-mode instability 
could be a possible source for gravitational radiation \cite{And01,Nar}. And, 
the
detection of gravitational emission due to r-mode instability may shed light on
the interior composition of a neutron star.

It was argued that the r-mode instability could
be effectively suppressed by bulk viscosity due to exotic matter in neutron 
star interior. Neutron star matter encompasses a wide range of densities, from 
the density of iron nucleus at the surface of the star to several times 
normal nuclear matter density in the core. Since the chemical potentials of
nucleons and leptons increase rapidly with density in the interior of neutron
stars, different kinds of exotic matter with large strangeness fraction such 
as, hyperon matter, Bose-Einstein condensed matter of antikaons 
and quark matter may appear there. The coefficient of bulk viscosity due to 
non-leptonic weak processes involving hyperons was calculated by several
authors \cite{Jon1,Jon2,Lin02,Dal,Dra}. It was noted that when the 
gravitational-radiation growth time scale was longer than the damping time 
scale due to hyperon bulk viscosity, the r-mode instability would be completely
suppressed.  Actually, a window of instability was found in the calculations of
Ref.\cite{Lin02,Dra}. Also, the impact of bulk viscosity due to unpaired and 
paired quark matter on the r-mode instability had been investigated extensively
\cite{Dra,Mad92,Mad00,Gon}.

Earlier studies showed that hyperons might appear in neutron star matter around 
2-3 normal nuclear matter density \cite{Bal,Ban}. It was also noted that 
$\Lambda$ hyperons appeared first in the system followed by $\Xi$ hyperons 
\cite{Ban}. On the other hand $\Sigma$ hyperons were excluded
from the system because of a repulsive $\Sigma$ hyperon potential depth at 
normal nuclear matter density \cite{Fri}. Also, hyperon-hyperon interaction
becomes important because the matter is hyperon-rich at high density regime.
This interaction was accounted by the exchange of two strange mesons, 
$f_0$ (975) (denoted 
hereafter as $\sigma^*$) and $\phi$ (1020) \cite{Sch93,Mis}. It was found that 
the competition between attractive $\sigma^*$ and repulsive $\phi$ fields made 
the overall equation of state stiffer at higher densities \cite{Mis}. In 
previous investigations of hyperon bulk viscosity, $\Sigma$ hyperons appeared 
before $\Lambda$ hyperons because an attractive  $\Sigma$ hyperon potential at 
normal nuclear matter density was considered. Those calculations also did not 
include hyperon-hyperon interaction.

In this paper, we investigate the effect of hyperon-hyperon interaction on bulk
viscosity coefficient, the corresponding damping timescale and the r-mode 
stability. Further, we use recent experimental data of hypernuclei to
determine hyperon-scalar meson couplings in this calculation. This paper is 
organized in the following way. In Sec. II, we describe the model to calculate
equation of state (EoS), bulk viscosity coefficient and the corresponding 
timescale. Parameters and results of our calculation are
discussed in Sec. III. We summarise and conclude in Sec. IV.

\section{Formalism}

We adopt a relativistic field theoretical model to describe the 
$\beta$-equilibrated and charge neutral hadronic matter. 
The constituents 
of matter are ${n, p, \Lambda, \Sigma^+, \Sigma^-, \Sigma^0, \Xi^-, \Xi^0 }$ of
the baryon octet and electrons and muons. In this model, 
baryon-baryon interaction is mediated by the exchange
of scalar and vector mesons and for hyperon-hyperon interaction, two 
additional mesons - scalar 
$\sigma^*$ and vector $\phi$ \cite{Sch93,Mis} are incorporated.
Therefore the Lagrangian density for the hadronic phase is given by
\begin{eqnarray}
{\cal L}_B &=& \sum_B \bar\Psi_{B}\left(i\gamma_\mu{\partial^\mu} - m_B
+ g_{\sigma B} \sigma - g_{\omega B} \gamma_\mu \omega^\mu
- g_{\rho B}
\gamma_\mu{\mbox{\boldmath t}}_B \cdot
{\mbox{\boldmath $\rho$}}^\mu \right)\Psi_B\nonumber\\
&& + \frac{1}{2}\left( \partial_\mu \sigma\partial^\mu \sigma
- m_\sigma^2 \sigma^2\right) - U(\sigma) \nonumber\\
&& -\frac{1}{4} \omega_{\mu\nu}\omega^{\mu\nu}
+\frac{1}{2}m_\omega^2 \omega_\mu \omega^\mu
- \frac{1}{4}{\mbox {\boldmath $\rho$}}_{\mu\nu} \cdot
{\mbox {\boldmath $\rho$}}^{\mu\nu}
+ \frac{1}{2}m_\rho^2 {\mbox {\boldmath $\rho$}}_\mu \cdot
{\mbox {\boldmath $\rho$}}^\mu + {\cal L}_{YY}~.
\end{eqnarray}
The isospin multiplets for baryons B $=$ N, $\Lambda$, $\Sigma$ and $\Xi$ are
represented by the Dirac spinor $\Psi_B$ with vacuum baryon mass $m_B$,
and isospin operator ${\mbox {\boldmath t}}_B$ and $\omega_{\mu\nu}$ and 
$\rho_{\mu\nu}$ are field strength tensors. The scalar
self-interaction term \cite{Bog} is
\begin{equation}
U(\sigma) = \frac{1}{3} g_2 \sigma^3 + \frac{1}{4} g_3 \sigma^4 ~.
\end{equation}
The Lagrangian density for hyperon-hyperon interaction (${\cal L}_{YY}$)
is given by
\begin{eqnarray}
{\cal L}_{YY} &=& \sum_B \bar\Psi_{B}\left(
g_{\sigma^* B} \sigma^* - g_{\phi B} \gamma_\mu \phi^\mu
\right)\Psi_B\nonumber\\
&& + \frac{1}{2}\left( \partial_\mu \sigma^*\partial^\mu \sigma^*
- m_{\sigma^*}^2 \sigma^{*2}\right)
-\frac{1}{4} \phi_{\mu\nu}\phi^{\mu\nu}
+\frac{1}{2}m_\phi^2 \phi_\mu \phi^\mu~.
\label{Lag}
\end{eqnarray}

We perform this calculation in the mean field approximation \cite{Ser}. The 
mean values for
corresponding meson fields are denoted by $\sigma$, $\sigma^*$, $\omega_0$, 
$\rho_{03}$ and $\phi_0$. Therefore, we replace meson fields with their 
expectation values and meson field equations become
\begin{equation}
m_{\sigma}^2\sigma=-{\partial U\over\partial\sigma}
+\sum_{B}g_{\sigma B}n_{B}^s~,
\end{equation}
\begin{equation}
m_{\sigma^*}^2\sigma^*=\sum_{B}g_{\sigma^* B}n_{B}^s~,
\end{equation}
\begin{equation}
m_{\omega}^2\omega_{0}=\sum_{B}g_{\omega B}n_{B}~,
\end{equation}
\begin{equation}
m_{\phi}^2\phi_{0}=\sum_{B}g_{\phi B}n_{B}~,
\end{equation}
\begin{equation}
m_{\rho}^2\rho_{03}=\sum_{B}g_{\rho B}I_{3B}n_{B}~.
\end{equation}
The scalar density and baryon number density are 
\begin{eqnarray}
n_B^S &=& \frac{2J_B+1}{2\pi^2} \int_0^{k_{F_B}} 
\frac{m_B^*}{(k^2 + m_B^{* 2})^{1/2}} k^2 \ dk ~,
\end{eqnarray}
\begin{eqnarray}
n_B &=& (2J_B+1)\frac{k^3_{F_B}}{6\pi^2} ~, 
\end{eqnarray}
where Fermi momentum is $k_{F_B}$, spin is $J_B$, and isospin projection is
$I_{3B}$. Effective mass and chemical potential of baryon $B$ are 
$m_B^*=m_B - g_{\sigma B}\sigma - g_{\sigma^* B}\sigma^*$ and
$\mu_{B} = (k^2_{F_{B}} + m_B^{* 2} )^{1/2} + g_{\omega B} \omega_0
+ g_{\phi B} \phi_0 + I_{3B} g_{\rho B} \rho_{03}$, respectively. 

As the hadronic phase is charge neutral, the total charge density is
\begin{equation}
Q = \sum_B q_B n_B -n_e -n_\mu =0~,
\end{equation}
where $n_B$ is the number density of baryon B, $q_B$ is the electric charge 
and $n_e$ and $ n_\mu$ are charge densities of electrons and muons respectively.

In compact star interior, hyperons maintain chemical equilibrium through weak 
processes. The generalised $\beta$-decay processes may be written in the form
$B_1 \longrightarrow B_2 + l+ \bar \nu_l$ and 
$B_2 +l \longrightarrow B_1 +\nu_l$ where $B_1$ and $B_2$ are baryons and 
l is a lepton. 
Therefore the generic equation for chemical equilibrium condition is
\begin{equation}
\mu_i = b_i \mu_n - q_i \mu_e ~,
\end{equation}
where $\mu_n$, $\mu_e$ and $\mu_i$ are respectively
the chemical potentials of neutrons, electrons and i-th baryon 
and $b_i$ and $q_i$ are baryon and electric charge of i-th baryon respectively.
The above equation implies that there are two independent chemical
potentials $\mu_n$ and $\mu_e$ corresponding to two conserved charges i.e. 
baryon number and electric charge.

The total energy density is given by 
\begin{eqnarray}
{\varepsilon}
&=& \frac{1}{2}m_\sigma^2 \sigma^2 
+ \frac{1}{3} g_2 \sigma^3 + \frac{1}{4} g_3 \sigma^4
+ \frac{1}{2}m_{\sigma^*}^2 \sigma^{*2}\nonumber\\ 
&& + \frac{1}{2} m_\omega^2 \omega_0^2 + \frac{1}{2} m_\phi^2 \phi_0^2 
+ \frac{1}{2} m_\rho^2 \rho_{03}^2 \nonumber\\
&& + \sum_B \frac{2J_B+1}{2\pi^2} 
\int_0^{k_{F_B}} (k^2+m^{* 2}_B)^{1/2} k^2 \ dk \nonumber\\
&& + \sum_l \frac{1}{\pi^2} \int_0^{K_{F_l}} (k^2+m^2_l)^{1/2} k^2 \ dk~.
\end{eqnarray}
And the pressure is 
\begin{eqnarray}
P &=& - \frac{1}{2}m_\sigma^2 \sigma^2 - \frac{1}{3} g_2 \sigma^3 
- \frac{1}{4} g_3 \sigma^4 \nonumber\\
&& - \frac{1}{2}m_{\sigma^*}^2 \sigma^{*2} 
+ \frac{1}{2} m_\omega^2 \omega_0^2 + \frac{1}{2} m_\phi^2 \phi_0^2 
+ \frac{1}{2} m_\rho^2 \rho_{03}^2 \nonumber\\
&& + \frac{1}{3}\sum_B \frac{2J_B+1}{2\pi^2} 
\int_0^{k_{F_B}} \frac{k^4 \ dk}{(k^2+m^{* 2}_B)^{1/2}}\nonumber\\
&& + \frac{1}{3} \sum_l \frac{1}{\pi^2} 
\int_0^{K_{F_l}} \frac{k^4 \ dk}{(k^2+m^2_l)^{1/2}}~. 
\end {eqnarray}

Bulk viscosity arises because pressure and density variations 
associated with the r-mode oscillation drive the system out of chemical
equilibrium. Various microscopic reaction processes bring the system back to 
equilibrium. Weak interaction processes are most important in this case. As we
are concerned about bulk viscosity coefficient in young neutron stars where 
temperature is $\sim$ 10$^9$ - 10$^{10}$ K, it was shown by various authors 
\cite{Nar,Jon1,Jon2,Lin02} that non-leptonic processes involving hyperons might
lead to a high value for the bulk viscosity coefficient. Though there may be 
several processes contributing to the bulk viscosity, only those reactions 
which might provide an upper limit on the bulk viscosity would be of interest. 
The relevant non-leptonic processes are 
\begin{eqnarray}
n + p \rightleftharpoons p + \Lambda~,\\
n + n \rightleftharpoons p + \Sigma~,\\
n + n \rightleftharpoons n + \Lambda~.
\end{eqnarray}
All these reactions involve variation of neutron number density ($n_n$) due to 
density perturbation on the system. Therefore, we consider neutron fraction
as a primary variable.

The real part of bulk viscosity coefficient is calculated in terms of 
relaxation times of microscopic processes \cite{Lin02,Lan} 
\begin{equation}
\zeta = \frac {P(\gamma_{\infty} - \gamma_0)\tau}{1 + {(\omega\tau)}^2}~,
\end{equation}
where $P$ is the pressure, $\tau$ is the net microscopic relaxation time and 
$\gamma_{\infty}$ and $\gamma_0$ are 'infinite' and 'zero' frequency adiabatic 
indices respectively. The factor
\begin{equation}
\gamma_{\infty} - \gamma_0 = - \frac {n_b^2}{P} \frac {\partial P} 
{\partial n_n} \frac {d{\bar x}_n} {dn_b}~,
\end{equation}
can be determined from the EoS. Here $\bar x_n = \frac {n_n}{n_b}$ gives the 
neutron fraction in the equilibrium state and $n_b = {\sum}_{B} n_B$ is the 
total baryon density. In the co-rotating frame, the angular velocity 
($\omega$) of (l,m) r-mode is 
related to angular velocity ($\Omega$) of a rotating 
neutron star as $\omega = {\frac {2m}{l(l+1)}} \Omega$ \cite{And01}. 

Here we consider the process in Eq. (15) because other processes may not 
contribute significantly to the calculation of bulk viscosity coefficient and 
are discussed
later in details. The relaxation time ($\tau$) for the process is given 
by \cite{Lin02}
\begin{equation}
\frac {1}{\tau} = \frac {{(kT)}^2}{192{\pi}^3} {p_{\Lambda}}
{<{{|M_{\Lambda}|}^2}>} \frac {\delta \mu}{{n_b}\delta{x_n}}~,
\end{equation}
where $p_{\Lambda}$ is the Fermi momentum for $\Lambda$ hyperons and 
$<{|M_{\Lambda}|}^2>$ is the angle averaged matrix element squared
given by Ref.\cite{Nar,Lin02}. Also, we have 
\begin{equation}
\frac {\delta \mu}{{n_b}\delta{x_n}} = {\alpha_{nn}} - {\alpha_{\Lambda n}}
- {\alpha_{n \Lambda}} + {\alpha_{\Lambda \Lambda}}~,
\end{equation}
where $\delta {x_n} = x_n - {\bar x}_n$ is the departure of neutron fraction
from its thermodynamic equilibrium value ${\bar x}_n$ and 
$\alpha_{ij} = {\frac {\partial\mu_i}{\partial n_j}}_{{n_k},k \neq j}$
which is determined from the EoS. As soon as we know the
relaxation time, we can calculate the bulk viscosity coefficient. 

The bulk viscosity damping timescale ($\tau_B$) due to the non-leptonic process 
involving hyperons is given by \cite{Nar,Lin02} 
\begin{equation}
{\frac {1}{\tau_B}} =  - {\frac {1} {2E}} {\frac {dE}{dt}}~,
\end{equation}
where E is the energy of the perturbation as measured in the co-rotating frame
of the fluid and is expressed as
\begin{equation}
E = \frac {1}{2}{\alpha^2}{\Omega^2}{R^{-2}} \int_0^R {\epsilon (r) r^6}dr~.
\end{equation}
Here, $\alpha$ is the dimensionless amplitude of the r-mode, R is the 
radius of the star and $\epsilon$(r) is the energy density profile. The 
derivative of the co-rotating frame energy with respect
to time is
\begin{equation}
\frac {dE}{dt} = -4 \pi \int_0^R \zeta (r) <|\vec{\nabla} \cdot 
{\delta \vec{v}}|^2> r^2 dr,~
\end{equation} 
where the angle average of the square of the hydrodynamic expansion 
\cite{Lin99} is $<|\vec{\nabla} \cdot {\delta \vec{v}}|^2> = \frac 
{({\alpha \Omega})^2}{690}
({\frac {r}{R}})^6 (1 + 0.86 ({\frac {r}{R}})^2) ({\frac {\Omega^2}
{\pi G \bar {\epsilon}}})^2$
and $\bar {\epsilon}$ is the mean energy density of a non-rotating star. 
Equation (24) is the bulk viscosity contribution to the imaginary 
part of the frequency of the r-mode. Taking into account time scales for 
gravitational radiation ($\tau_{GR}$) and modified Urca process ($\tau_U$) 
involving only nucleons, we can define the overall r-mode time scale ($\tau_r$)
as
\begin{equation}
{\frac {1}{\tau_r}} =  - {\frac {1}{\tau_{GR}}} + {\frac {1}{\tau_B}} + 
{\frac {1}{\tau_U}}.~ 
\end{equation}
Here $\frac {1}{\tau_r}$ is the imaginary part of the frequency of the r-mode.
The gravitational radiation timescale is given by \cite{Lin98}
\begin{equation}
{\frac {1}{\tau_{GR}}} =  
\frac {131072 \pi}{164025} {\Omega^6} \int_0^R {\epsilon (r) r^6}dr~.
\end{equation}
The time scale due to modified Urca process ($\tau_U$) involving only nucleons 
is calculated similarly as the hyperon bulk viscosity damping time scale but
using the following expression for bulk viscosity coefficient for modified Urca 
process given by \cite{Lin98,Saw}
\begin{equation}
\zeta_U = 6 \times 10^{-59} \epsilon^2 T^6 \omega^2~.
\end{equation}

Damping timescales are functions of angular velocity, neutron star mass
and temperature. 
Therefore, solving $\frac {1}{\tau_r}$ = 0, we calculate the critical 
angular velocity above which the r-mode is unstable whereas it is stable below 
the critical angular velocity. 

\section{Results and Discussion}

In this calculation, we require the knowledge of coupling constants for baryons
with $\sigma$, $\omega$, $\rho$, $\sigma^*$ and $\phi$ mesons. Nucleon-meson
coupling constants are obtained by reproducing nuclear matter saturation
properties and taken from Ref. \cite{Gle91}. Here we consider the parameter set
corresponding to the 
value of incompressibility of nuclear matter K = 240 MeV.
Nucleon-meson coupling constants are listed in Table I. 

Hyperon-vector meson coupling constants are determined from SU(6) symmetry of
the quark model \cite{Mis,Dov,Sch94}. The scalar $\sigma$ meson 
coupling to hyperons is calculated from the potential depth of a hyperon (Y) 
\begin{equation}
U_Y^N (n_0) = - g_{\sigma Y} \sigma + g_{\omega Y} \omega_0~,
\end{equation}
in normal nuclear matter. The potential depth of $\Lambda$ hyperons in normal
nuclear matter $U_{\Lambda}^N (n_0)$ = -30 MeV is obtained from the analysis
of energy levels of $\Lambda$ hypernuclei \cite{Dov,Chr}. Recent 
$\Xi$-hypernuclei data from various experiments \cite{Fuk,Kha} give 
a relativistic potential of $U_{\Xi}^N (n_0)$ = -18 MeV. However, the analysis 
of $\Sigma^-$ 
atomic data implies a strong isoscalar repulsion for $\Sigma^-$ hyperons in
nuclear matter \cite{Fri}. Also, recent $\Sigma$ hypernuclei data indicate a 
repulsive
$\Sigma$-nucleus potential depth \cite{Bart}. Therefore, we adopt a repulsive
potential depth of 30 MeV for $\Sigma$ hyperons \cite{Fri}. 

The hyperon-$\sigma^*$ coupling constants are estimated by fitting them to a
potential depth, ${U_{Y}^{(Y^{'})}}{(n_0)}$, for a hyperon (Y) in a hyperon 
($Y^{'}$) matter at normal nuclear matter density obtained from double 
$\Lambda$ hypernuclei data \cite{Sch93,Mis}. This is given by
\begin{equation}
U_{\Xi}^{(\Xi)}(n_0) = U_{\Lambda}^{(\Xi)}(n_0) = 2 U_{\Xi}^{(\Lambda)}(n_0)
= 2 U_{\Lambda}^{(\Lambda)}(n_0) = -40~.
\end{equation}

Now we perform our calculation employing nucleon-meson and hyperon-meson
coupling constants corresponding to K = 240 MeV case. We find that before
the appearance of hyperons, the $\beta$-equilibrated matter is composed of
neutrons, protons, electrons and muons and their abundances increase with
baryon density. Here, charge neutrality is maintained by protons, electrons and
muons. Particle populations as a function of baryon density and without 
hyperon-hyperon interaction are displayed in Fig. 1. 
At 2.6$n_0$, $\Lambda$ hyperons appear first in the system. It is 
observed that the density of $\Lambda$ hyperon increases whereas the neutron 
density decreases with increasing baryon density for the growth of $\Lambda$ 
hyperons occurs at the expense of neutrons. Next $\Xi^-$ hyperons are populated
around 3.0$n_0$. However, $\Sigma$ hyperons do not appear in the system even at
higher densities because of repulsive $\Sigma$-nuclear matter interaction. 
Further we note that threshold densities for the appearance of different 
hyperon species except $\Xi^0$ hyperons are not changed appreciably due to 
hyperon-hyperon interaction and this is demonstrated in Fig. 2.

Equations of state (pressure versus energy density) for different compositions
are shown in Fig. 3. The bold solid curve implies nucleons-only matter whereas
the dashed and solid curves correspond to hyperon matter with and without
interaction, respectively. The appearance of hyperons makes equations of state
softer. On the other hand, the EoS with hyperon-hyperon interaction is initially
softer compared with the case without the interaction, but it becomes stiffer
later. This may be understood in the following way. Hyperon-hyperon interaction
is mediated by $\sigma^*$ and $\phi$ mesons. The additional attraction of 
$\sigma^*$ field makes the EoS softer. The repulsive contribution of $\phi$ 
field becomes dominant at higher densities. There is a competition between these
two effects. Consequently, the EoS changes over from softer to stiffer at 
higher energy densities.   

Now we calculate the bulk viscosity coefficient using the EoS 
including hyperon-hyperon interaction. Already, we have noted that $\Lambda$ 
hyperons are populated first in the system and heavier hyperons appear at 
higher densities. Therefore, we are interested in non-leptonic processes, 
Eqs. (15) and (17), involving $\Lambda$ hyperons. However, it was 
noted that transition rate for the non-leptonic process in Eq. (17) was
one order of magnitude greater than that   of the process in Eq. (15) 
and led to a lower relaxation time in the former case \cite{Jon2}. Therefore,
we only consider the process in Eq. (15). In this
calculation, partial derivatives of pressure and chemical potentials with
respect to particle number density are determined with the help of the EoS.
In Fig. 4, the difference of fast and slow adiabatic indices ($\gamma_{\infty} -
\gamma_0$) is plotted with normalised
baryon density. The dashed curve represents the case with hyperon-hyperon
interaction whereas the solid curve indicates the case without the 
interaction. The dashed curve rises to a maximum value and then
drops below the solid curve. This can be attributed to the interplay
of $\sigma^*$ and $\phi$ fields and the crossover from a softer to a stiffer
EoS.  

Next we compute the relaxation time as given by Eq. (20) for the process in 
Eq. (15). Effective masses, Fermi momenta, chemical potentials of baryons and
partial derivatives entering into the expression of the relaxation time are 
obtained from the EoS. In the calculation of matrix element, we use the values
of axial-vector coupling constants $g_{np}$ = -1.27 and $g_{p\Lambda}$ = -0.72
obtained from $\beta$-decay of baryons at rest. We exhibit the relaxation time 
as a function of normalised baryon density in Fig. 5. The temperature is set to
$10^{10}$ K in this case. The dashed and solid curves denote
cases with and without hyperon-hyperon interaction, respectively. We observe
that the effect of the interaction on the relaxation time is not significant. 
Further, the behaviour of relaxation time with temperature for hyperon-hyperon
interaction case is shown in Fig. 6.
The relaxation time increases as the temperature is decreased. The effect of
temperature on the relaxation time is substantial.

The hyperon bulk viscosity coefficient is plotted with normalised baryon 
density for a temperature $10^{10}$ K in Fig. 7. For this calculation we 
obtain angular velocities
($\Omega$) at different central densities corresponding to a sequence of 
neutron stars using the rotating star model \cite{Ster95,Note} and equations of 
state with and without hyperon-hyperon interaction. We note that the factor 
$\omega \tau$ is much much less than 1 over the range of
baryon densities that we have shown here and is neglected in the bulk viscosity
coefficient given by Eq. (18). The bulk viscosity coefficient in 
hyperon-hyperon interaction case represented by the dashed curve becomes smaller
than that without the interaction denoted by the solid curve in the 
density regime which might occur in maximum mass neutron stars. The effect of
the interaction enters into the bulk viscosity coefficient through the pressure
term and adiabatic indices. The bulk viscosity coefficient increases appreciably
with decreasing temperature as it is evident from Fig. 8. The large value of 
$\zeta$ at $10^9$ K indicates that the viscous damping might be important in 
suppressing r-mode instability in neutron stars. It is to be noted 
here that we have neglected $\Lambda$-hyperon superfluidity in this 
calculation. Earlier calculations of bulk viscosity with hyperons involved
hyperon superfluidity \cite{Nar,Lin02}. However, it has been reported that 
superfluid gaps for hyperons would be very small \cite{Tak,Tsu}. 

Now we discuss the effect of the EoS including hyperon-hyperon interaction on
the stability of r-mode. This mode becomes stable when the gravitational 
radiation timescale is greater than the damping timescales due to hyperon bulk 
viscosity as well as modified Urca bulk viscosity. The calculation of bulk
viscosity timescales as given by Eqs. (22)-(24) needs the structure of a 
neutron star and its energy
density profile. We have calculated the sequences of non-rotating neutron
stars using Tolman-Oppenheimer-Volkoff equations with and without 
hyperon-hyperon 
interaction. The maximum neutron star masses with and without hyperon-hyperon 
interaction are 1.64 M$_{solar}$ and 1.69 M$_{solar}$ corresponding to central 
baryon densities 7.1 and 6.9 $n_0$, respectively. Similarly, we have computed 
sequences of rotating stars with and without hyperon-hyperon interaction using
rotating neutron star model of Stergioulas \cite{Ster,Ster95,Note}. In this 
case, maximum masses of rotating neutron stars with and without hyperon-hyperon 
interaction are 1.95 M$_{solar}$ and 2.00 M$_{solar}$ corresponding to central
baryon densities 5.4 and 5.7 n$_0$ respectively. 

Next we perform the calculation of damping timescale due to hyperon bulk 
viscosity with and without hyperon-hyperon interaction for rotating neutron 
stars of mass 1.6 M$_{solar}$. This requires the knowledge of energy density
profiles in rotating neutron stars. Earlier investigations adopted slowly
rotating neutron stars and used the energy density profiles of non-rotating
neutron stars \cite{Lin02}. Recently, Nayyar and Owen \cite{Nar} incorporated 
the effects of rotation in the calculation of critical angular velocity using 
Hartle's slow rotation approximation. Here we also include the effects of 
rotation on damping time scale using energy density profiles of rotating 
neutron stars. A rotating neutron star has less central energy density 
than that of its non-rotating counterpart having the same gravitational mass
because of the contribution from centrifugal force. Therefore, hyperon 
population will be sensitive to the rotation. Here we are dealing with a 
situation where the young neutron star is rotating very fast and radiating 
gravitational waves giving rise to the r-mode instability in it. As the neutron 
star emits gravitational radiation it spins down. Consequently, the central 
density in it increases. When the baryon density in the interior is such that 
hyperon thresholds are reached, hyperons will be produced abundantly. In this 
calculation, we consider 1.6 M$_{solar}$ mass neutron stars with and without 
hyperon-hyperon interaction having 
central baryon densities 3.9 and 3.6 $n_0$ which are much above the threshold 
for $\Lambda$ hyperons and rotating at  angular velocities 
$\Omega_{rot}$ = 2952 $s^{-1}$ and 3100 $s^{-1}$  respectively.
We calculate the hyperon bulk
viscosity damping time scales using the energy density profiles of those stars.
Their energy density profiles with and without 
hyperon-hyperon interaction are displayed in Fig. 9. The dashed curve including
the interaction has larger energy densities than the case without the 
interaction. It is to be noted that  
the Keplerian 
angular velocities of neutron stars of mass 1.6 M$_{solar}$ with and without 
the interaction are 5600 s$^{-1}$ and corresponding central baryon densities 
are below the hyperon thresholds. 
The angular velocity for l=m=2 r-mode for a neutron 
star rotating with angular velocity ($\Omega$) is given by 
$\omega = \frac {2}{3} \Omega$. In this calculation we obtain a set of values
for $\omega$ corresponding to $\Omega$ ranging from 0 to $\Omega_{rot}$. 
Also, we express hyperon bulk viscosity $\zeta$ as a 
function of $r$ using the knowledge of baryon density profiles. All these 
inputs are now used to calculate the damping timescale. Similarly, we 
calculate the damping timescale due to modified Urca process for nucleons using
the expression for the bulk viscosity coefficient Eq. (27). 
However, the bulk viscosity due to leptonic processes is several orders of
magnitude lower than that of non-leptonic processes and has small effect on the
damping of the mode. It is to be noted here that the effect of direct Urca 
process involving hyperons on bulk viscosity was investigated and it was found
to have a small effect on the damping of the r-mode \cite{Lin02}. As the
gravitational radiation drives the r-mode unstable, it comes with a 
negative sign in Eq. (25). It is worth 
mentioning here that the energy density profiles of rotating stars used in this
calculation do not differ much from their non-rotating counterparts.

Now we determine critical angular velocities ($\Omega_C$) as a function of 
temperature for a neutron star of mass 1.6 M$_{solar}$ by solving 
$\frac {1}{\tau_r}$ = 0 as given by Eq. (25). The
critical angular velocity is plotted with temperature in Fig. 10 for cases
with (dashed curve) and without (solid curve) hyperon-hyperon 
interaction. We find that two curves do differ a little bit at and above 
5 $\times 10^{9}$ K. This implies that gravitational radiation dominates in 
this region making the r-modes unstable. However, below this temperature 
damping timescales due to hyperon bulk viscosity starts dominating. 
The hyperon-hyperon interaction suppresses the instability more effectively 
below 5 $\times 10^{9}$ K. Consequently, the instability window shrinks. 

\section{Summary and conclusions}

We have studied the effect of exotic matter, in particular, hyperon matter 
including hyperon-hyperon interaction on bulk viscosity. Here we have 
constructed equations of state within 
the framework of a relativistic field theoretical model. As large number of 
hyperons may be produced in dense matter, hyperon-hyperon interaction becomes
important and have been included in our calculation. This interaction is 
mediated by two strange mesons. Here, we use recent 
hypernuclei data which give rise to attractive potential depths for $\Lambda$
and $\Xi$ hyperons and a repulsive potential depth for $\Sigma$ hyperons in
normal nuclear matter. Also, we exploit the knowledge of double $\Lambda$
hypernuclei data to find the strength of hyperon-$\sigma^*$ meson coupling
constant. Using these potential depths, we find that $\Lambda$
hyperons appears first in the system followed by $\Xi^-$ and $\Xi^0$ 
hyperons. However, $\Sigma$ hyperons do not appear because of the repulsive
potential. Hyperon-hyperon interaction 
makes the EoS softer resulting in a smaller maximum mass neutron star than
that without the interaction. Next we have computed the bulk
viscosity coefficient due to the non-leptonic weak process 
$n + p \rightleftharpoons p + \Lambda$ and its influence on r-mode stability.
We have used energy density profiles for rotating neutron stars to take into
account the effects of rotation on hyperon populations.
It is found that the gravitational radiation driven r-mode instability 
is more effectively suppressed due to the bulk viscosity coefficient
in  hyperon-hyperon interaction case compared with the situation without the 
interaction. 

Besides hyperon matter, there might be other forms of matter such as 
Bose-Einstein
condensates of antikaons and quarks. It was shown in earlier calculations that
Bose-Einstein condensation of antikaons might appear around 2-3$n_0$ and delay
the appearance of hyperons \cite{Bani}. So far, there is no calculation of bulk
viscosity in the antikaon condensed phase due to non-leptonic processes and how
it competes with hyperon bulk viscosity coefficient. This problem is being 
investigated by us and will be reported in a future publication.

{
  
} 

%=============================TABLE I ======================================
\begin{table}

\caption{The nucleon-meson coupling constants in
the GM1 set are taken from Ref. [29]. 
The coupling constants are obtained by reproducing the nuclear matter 
properties of binding energy $E/B=-16.3$ MeV, baryon density $n_0=0.153$ 
fm$^{-3}$, asymmetry energy coefficient $a_{\rm asy}=32.5$ MeV along with 
incompressibility $K=240$ MeV, and effective nucleon mass $m^*_N/m_N = 0.78$.
The hadronic masses are $m_N=938$ MeV, 
$m_\sigma=550$ MeV, $m_\omega=783$ MeV, and $m_\rho=770$ MeV. 
All coupling constants are dimensionless, except $g_2$ which is in fm$^{-1}$.}

\vspace{1.0cm}

\begin{tabular}{cccccc} 
\hline
\hline
K (MeV) & $g_{\sigma N}$& $g_{\omega N}$& $g_{\rho N}$&
$g_2$& $g_3$ \\ \hline
240& 8.7822&  8.7122& 8.5416& 27.8812& -14.3970 \\
\\
\hline
\hline
\\

\end{tabular}
\end{table}
\newpage
%%%%%%%%%%%%%%%%%%%%%%%%%%%%%%%%%%%%%%%%%%%%%%%%%%%%%%%%%%%%%%%%%%%%%%
\vspace{-2.0cm}

{\centerline{
\epsfxsize=12cm
\epsfysize=14cm
\epsffile{fig1.eps}
}}

\vspace{1.0cm}

\noindent{\small{
Fig. 1. Particles abundances are plotted with normalised baryon density for the
case without hyperon-hyperon interaction.}}

\newpage
\vspace{-2cm}

{\centerline{
\epsfxsize=12cm
\epsfysize=14cm
\epsffile{fig2.eps}
}}

\vspace{4.0cm}

\noindent{\small{
Fig. 2. Same as in Fig. 1 but for  hyperon-hyperon interaction case.}}

\newpage
\vspace{-2cm}

{\centerline{
\epsfxsize=12cm
\epsfysize=14cm
\epsffile{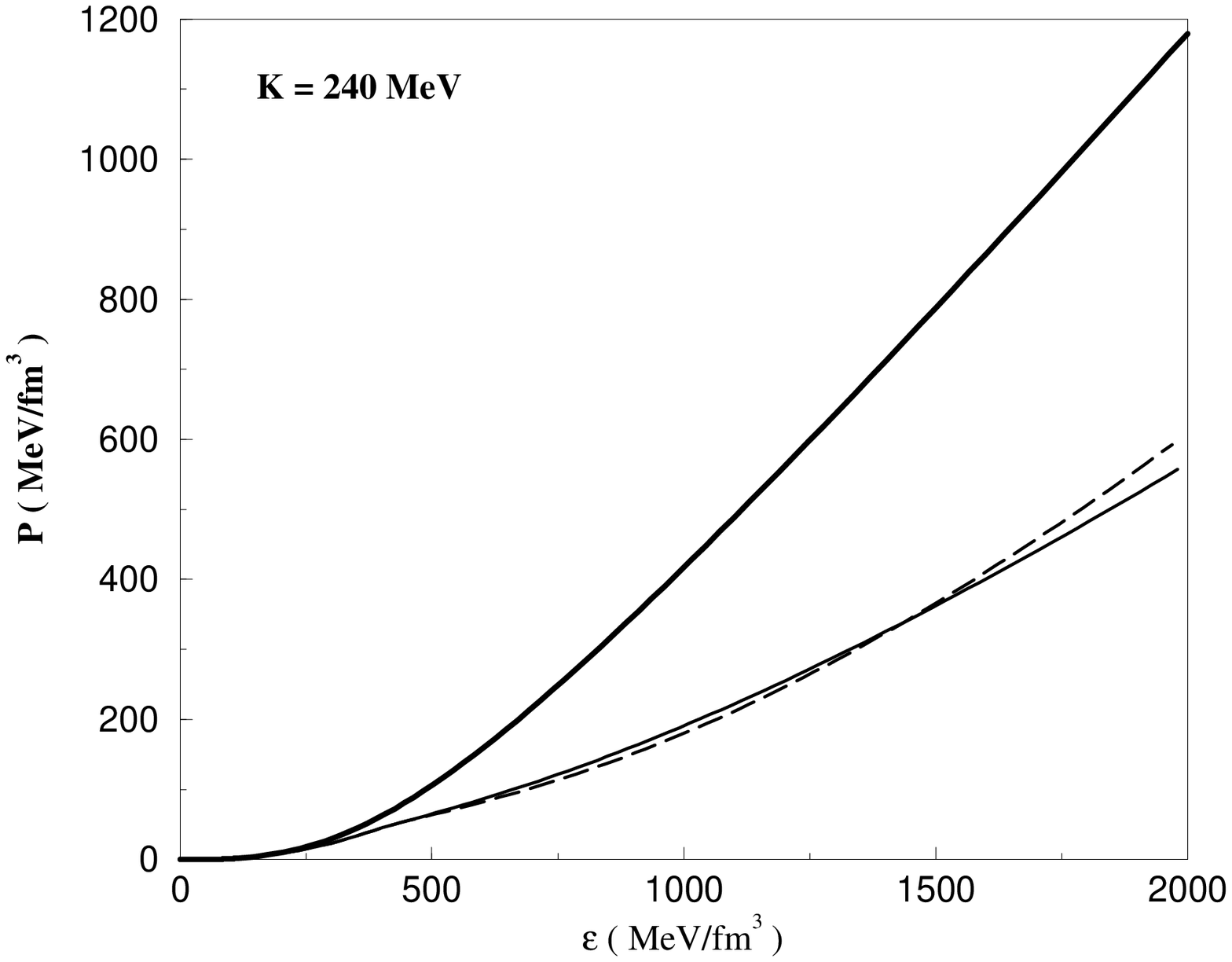}
}}

\vspace{4.0cm}

\noindent{\small{
Fig. 3. The equation of state, pressure P vs energy density $\epsilon$ , for
nucleons-only matter (bold solid curve), hyperon matter with (dashed curve) and
without (solid curve) hyperon-hyperon interaction are shown here.}}

\newpage
\vspace{-2cm}

{\centerline{
\epsfxsize=12cm
\epsfysize=14cm
\epsffile{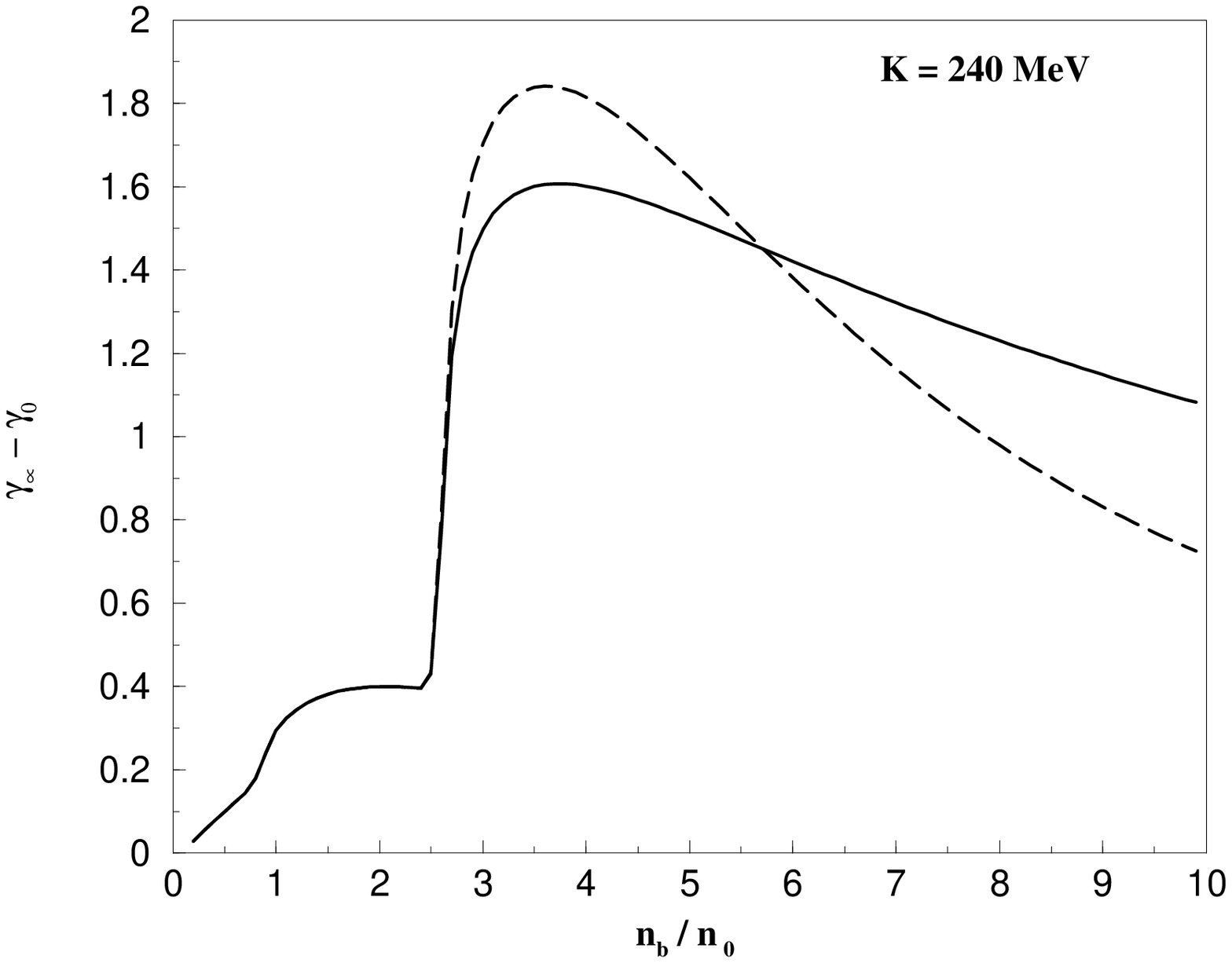}
}}

\vspace{4.0cm}

\noindent{\small{
Fig. 4. The difference of adiabatic indices, $\gamma_{\infty} - \gamma_0$, is
shown as a function of normalised baryon density with (dashed) and without 
(solid curve) hyperon-hyperon interaction.}}

\newpage
\vspace{-2cm}

{\centerline{
\epsfxsize=12cm
\epsfysize=14cm
\epsffile{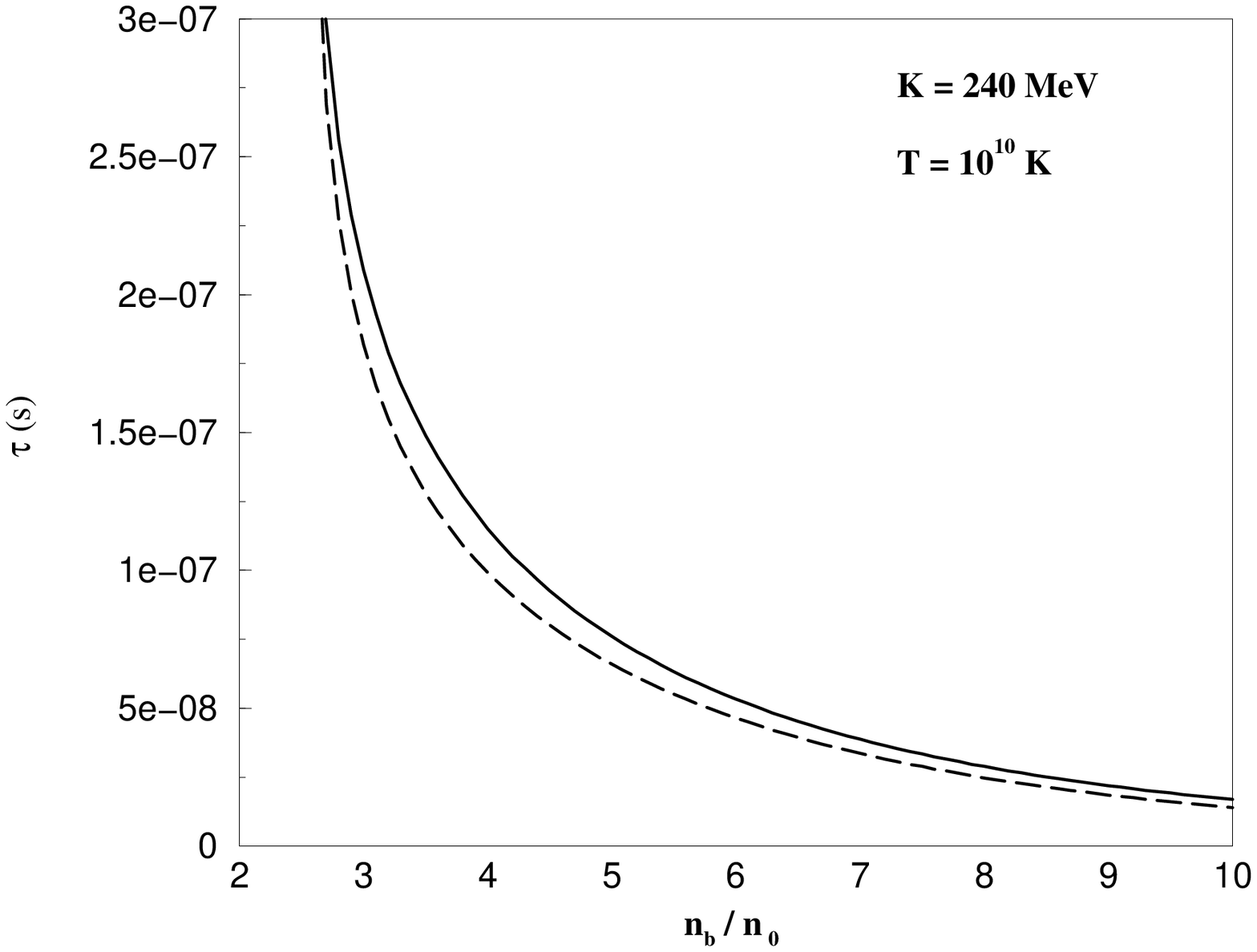}
}}

\vspace{4.0cm}

\noindent{\small{
Fig. 5. Relaxation time is plotted with normalised baryon density for the 
non-leptonic process in Eq. (15) at a temperature $10^{10}$ K for hyperon
matter with (dashed curve) and without (solid curve) hyperon-hyperon 
interaction.}}

\newpage
\vspace{-2cm}

{\centerline{
\epsfxsize=14cm
\epsfysize=12cm
\epsffile{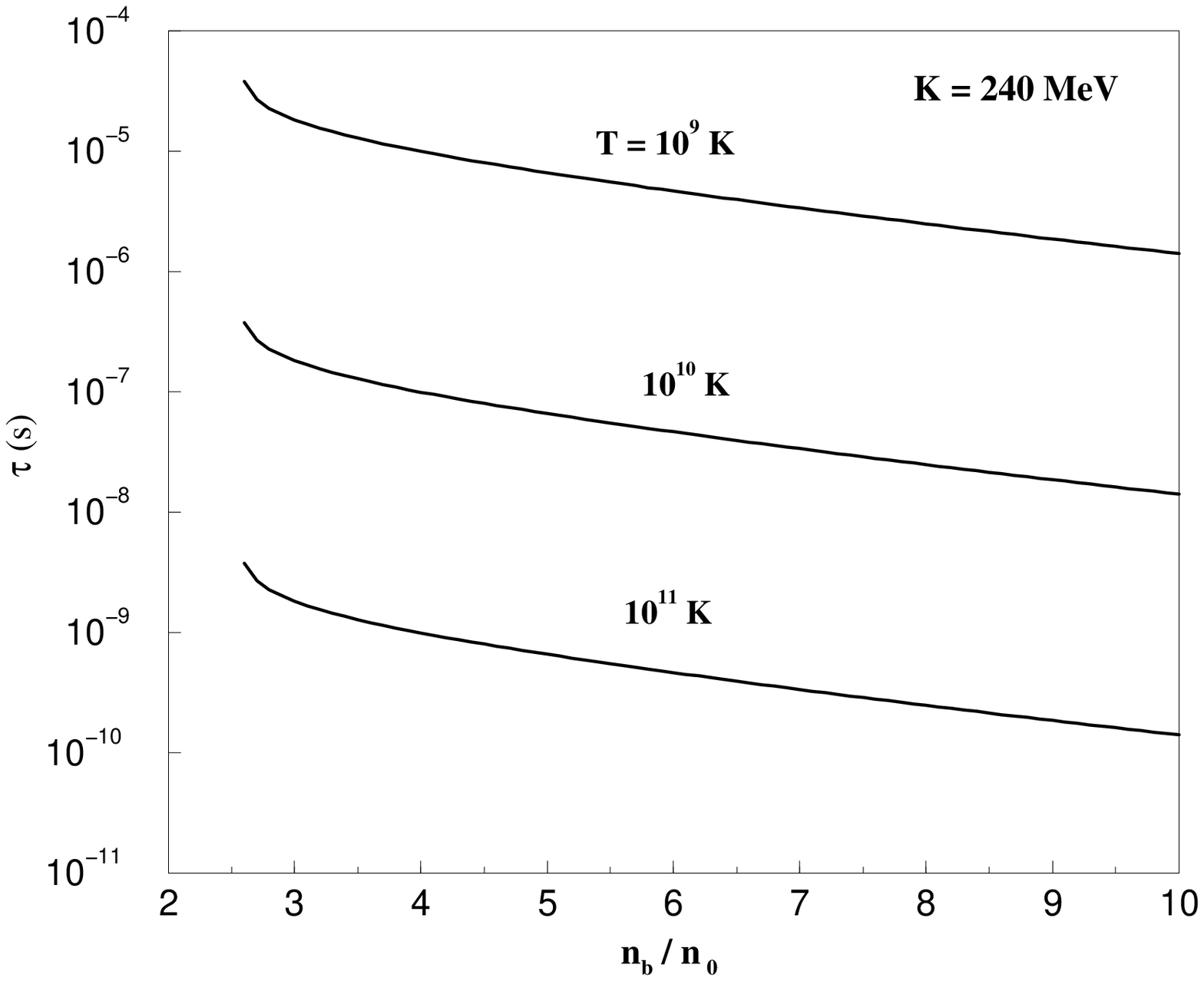}
}}

\vspace{4.0cm}

\noindent{\small{ Fig. 6. Same as in Fig. 5 but for different temperatures and
with hyperon-hyperon interaction.}}

\newpage
\vspace{-2cm}

{\centerline{
\epsfxsize=14cm
\epsfysize=12cm
\epsffile{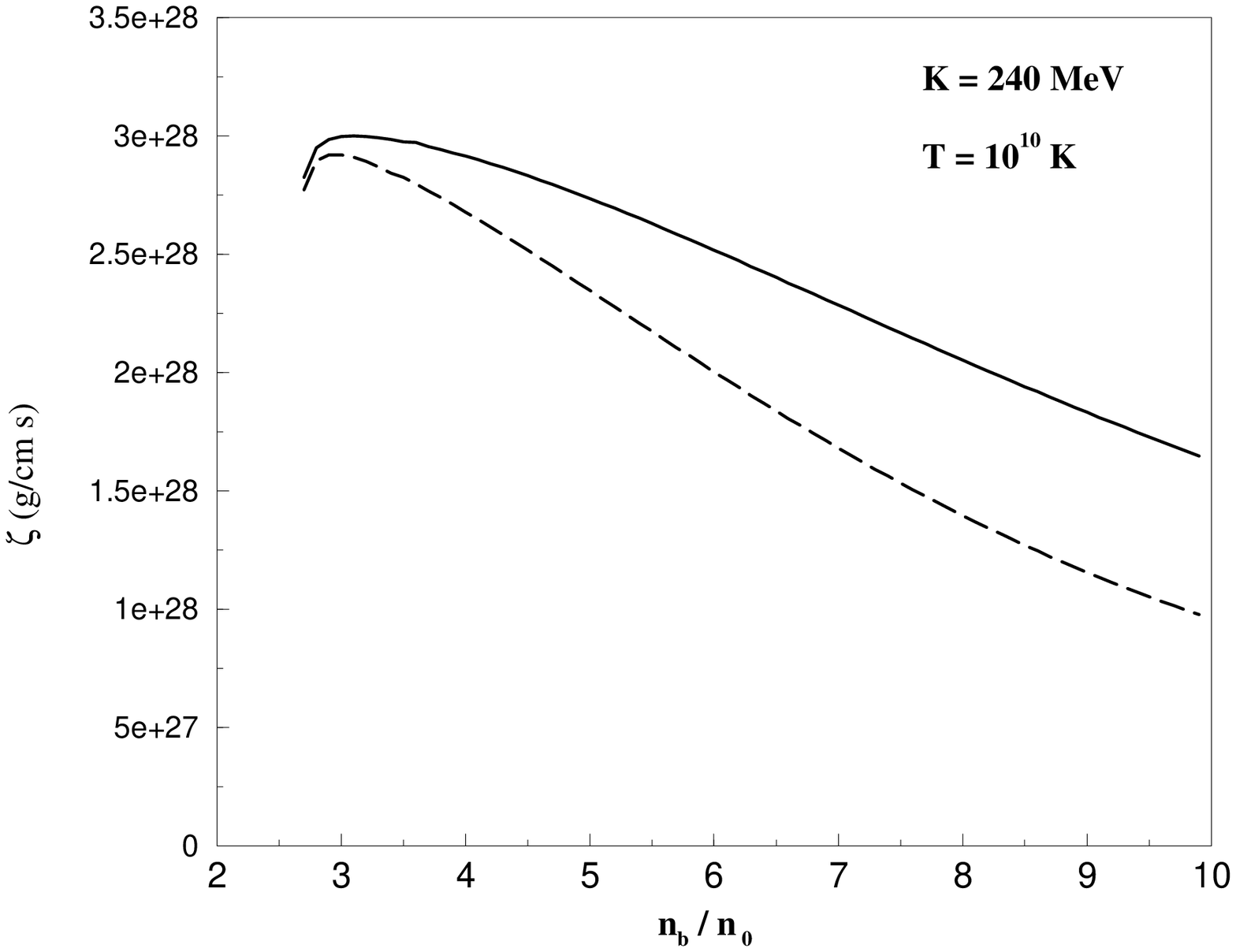}
}}

\vspace{4.0cm}

\noindent{\small{
Fig. 7. Bulk viscosity coefficient is exhibited as a function of normalised
baryon density for the process in Eq. (15) at a temperature $10^{10}$ K for
hyperon matter with (dashed curve) and without (solid curve) hyperon-hyperon 
interaction.}}

\newpage
\vspace{-2cm}

{\centerline{
\epsfxsize=14cm
\epsfysize=12cm
\epsffile{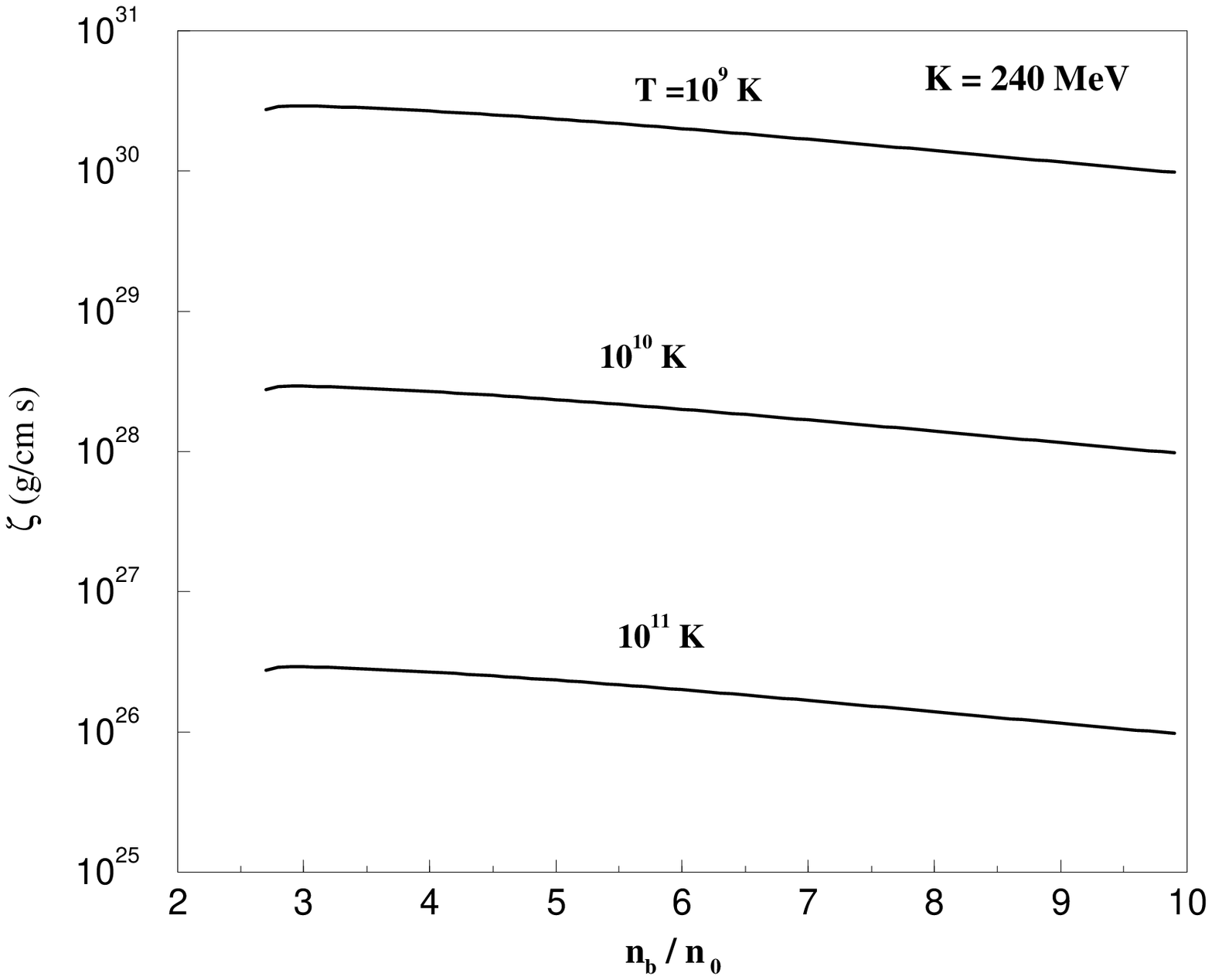}
}}

\vspace{4.0cm}

\noindent{\small{ Fig. 8. Same as in Fig. 5 but for different temperatures and
hyperon-hyperon interaction.}}

\newpage
\vspace{-2cm}

{\centerline{
\epsfxsize=14cm
\epsfysize=12cm
\epsffile{fig9.eps}
}}

\vspace{4.0cm}

\noindent{\small{
Fig. 9. Energy density profile is shown with radial distance for
rotating neutron stars of mass 1.6 M$_{solar}$ with (dashed curve) and
without (solid curve) hyperon-hyperon interaction.}}

\newpage
\vspace{-2cm}

{\centerline{
\epsfxsize=14cm
\epsfysize=12cm
\epsffile{fig10.eps}
}}

\vspace{4.0cm}

\noindent{\small{
Fig. 10. Critical angular velocities for 1.6 M$_{solar}$ neutron star are
plotted as a function of temperature with (dashed curve) and without 
(solid curve) hyperon-hyperon interaction.}}


\begin{thebibliography}{99}
\bibitem {Note} We adopt the RNS code of Stergioulas (Ref. [36]) to 
calculate the structure of rotating stars and their Kepler frequencies.
\end{thebibliography}
\end{document}